# First-principles pressure dependent investigation of the physical properties of $KB_2H_8$: a prospective high-$T_C$ superconductor

Md. Ashraful Alam[1,2], F. Parvin[2], S. H. Naqib[2]*

[1]Department of Physics, Mawlana Bhashani Science and Technology University, Santosh, Tangail 1902, Bangladesh

[2]Department of Physics, University of Rajshahi, Rajshahi 6205, Bangladesh

*Corresponding author; Email: *salehnaqib@yahoo.com*

**Abstract**

Using the density functional theory (DFT) based first-principles investigation, the structural, mechanical, hardness, elastic anisotropy, optoelectronic, and thermal properties of cubic $KB_2H_8$ have been studied within the uniform pressure range of 0 - 24 GPa. The calculated structural parameters are in good agreement with the previous theoretical work. The compound $KB_2H_8$ is found to be structurally and thermodynamically stable in the pressure range from 8 GPa to 24 GPa. Single crystal elastic constants $C_{ij}$ and bulk elastic moduli ($B$, $G$ and $Y$) increase systematically with pressure from 8 GPa to 24 GPa. In the stable phase, $KB_2H_8$ is moderately elastically anisotropic and ductile in nature. The compound is highly machinable and fracture resistant. The Debye temperature, melting temperature and thermal conductivity increases with pressure. The results of electronic band structure calculations and optical parameters at different pressures are consistent with each other. The compound is optically isotropic. The compound $KB_2H_8$ has potential to be used as a very efficient solar energy reflector. The electronic energy density of states at the Fermi level decreases systematically with increasing pressure. The same trend is found for the repulsive Coulomb pseudopotential. Possible relevance of the studied properties to superconductivity has also been discussed in this paper.

## 1. Introduction

In 2004, N. W. Ashcroft suggested high temperature superconductivity at room temperature through chemical pressure or external pressure in hydride materials [1]. Now a day, it is a big physical and technical challenge to get superconductivity in metallic hydrides at elevated temperature and reduced external pressure. One of the possibility is the of metallic modification of hydrogen containing compounds [2]. To get hydrides as a metal, one often needs high pressure and it is a very difficult technological challenge [3,4]. In addition to possible high-temperature superconductivity, there are various other applications of hydrides. For example, as hydrogen storage systems like metal (M) hydrides ($MH_n$) used for off-board storage, and hydrides used as cryo-coolers [5]. The binary compound $H_3S$ is a remarkable high-temperature superconductor with a superconducting $T_C$ of 203 K at 200 GPa [1,6]. The central physical feature of these superconducting hydrides is to be found within the energy scales of energies associated with the lattice dynamics. The vibrational energies are very high for the light atoms (hydrogen) but there are also lower energy phonon branches for the more massive atoms with much higher electronic charges leading to strong electron-phonon couplings. Some recent examples of binary hydrides with high predicted superconducting transition temperature, $T_C$, under pressure, are: $AcH_5$ [7] ($T_C$ ~ 79 K @ 150 GPa), $AcH_{10}$ and $UH_8$ [8] ($T_C$ ~ 204 K @ 200 GPa and $T_C$ ~ 193 K under ambient condition), $YH_{10}$ and $LaH_{10}$ [9] ($T_C$ ~ 326 K @ 250 GPa and $T_C$ ~ 286 K @ 220 GPa, respectively), $ThH_{10}$ [10] ($T_C$ ~ 221 K @ 100 GPa). Besides, Elliot et al. [11] reported superconducting transition temperature $T_C$ ~ 288 K in a carbonaceous sulfur hydride at 267 GPa. On the other hand, there are some ternary hydrides with high superconducting critical temperature under pressure such as $ScCaH_8$ ($T_C$ ~ 212 K @ 200 GPa) and $ScCaH_{12}$ ($T_C$ ~ 182 K @ 200 GPa) [12], $ScYH_6$ with $T_C$ ~ 32.11 K to 52.91 K in the pressure range 0 - 200 GPa [13], and $H_3SXe$ with a $T_C$ of 89 K at 240 GPa [14]. A few of these reported compounds have been synthesized; the others are predicted.

As mentioned above, the exploration of physical properties of hydride superconductors is experimentally challenging under high pressure. Phonon mediated superconductors depend primarily on two factors: (i) the strength of the electron-phonon coupling (EPC) and (ii) the characteristic energy scale of phonon excitations [15,16]. There are two guiding rules that can be used for hunting high-$T_C$ superconductivity in hydrogen-rich compounds under high pressure: (i) the existence of metallic hydrogen-related σ-bonding bands in the electron dispersion and (ii) a large electronic energy density of states contributed by the hydrogen orbitals around the Fermi level [16]. It is well known that the electronegativity of hydrogen is small in comparison to other atoms with comparatively large atomic number. For this reason, the 1$s$ orbital of H atom contributes to metallic conduction at ultrahigh pressures [17,18]. The very high pressure realization of metallic hydrides with potential high-temperature superconductivity poses serious problem for practical applications. Therefore, in recent times, researchers are trying to find out low pressure hydrogen-rich high-$T_C$ superconductors. It has been proposed that relatively low electronegativity, nonmetallic and low atomic mass element would be ideal bonding partner for hydrogen [19] at relatively lower pressure. Gao et al. [19] proposed $BH_4$ tetrahedron as the building block to produce different hydride superconductors such as $Zr(BH_4)_4$ [20]. To produce a stable and metallic compound Gao et al. [19] also introduced potassium (K) with $BH_4$ and as a result suggested the $KB_2H_8$ compound. Previous study on this compound focused on the electronic band structure, lattice dynamics and ECP only [19]. Furthermore, Gao et al. [19] found that $KB_2H_8$ is thermodynamically stable at 12 GPa and at higher pressures. Most of the pressure dependent physical properties of $KB_2H_8$ remain unexplored till now. The unexplored physical properties include structural, elastic, mechanical and bonding characteristics, Fermi surface features at different pressures, thermophysical parameters, and optical parameters. We have explored all these so far unexplored physical features in details in this study. The electronic band structures have been revisited. The novel results obtained in this work shed light on possible applications of $KB_2H_8$ and on its prospective pressure dependent high-$T_C$ superconductivity.

The rest of this work has been structured as follows. In section 2 we have briefly described the computational methodology. The results are presented and analyzed in section 3. Finally, the main conclusions of this study are summarized in section 4.

**2. Computational scheme**

For geometry optimization of the compounds of interest, plane wave pseudopotential [21,22] method was used which is implemented in the CASTEP code [21]. For the electronic exchange-correlation, generalized gradient approximation (GGA) with PBE functional [23] was used. For the calculations of electron-ion interactions ultrasoft pseudopotential was used [24]. The BFGS algorithm was used to minimize the total energy and internal forces [25] within the optimized crystal structure. The Monkhorst–Pack grid [26] was used for *k*-point sampling. The elastic constants were determined using the stress-strain module in CASTEP. Thermophysical parameters were computed from the elastic constants and moduli [27-29]. The optical constants were obtained from the electronic band structure via the matrix element of photon induced interband electronic transition. The details regarding this procedure can be found elsewhere [30-32]. All the DFT calculations were performed in the ultrafine mode. The tolerance levels and parameter settings used during computations are summarized below.

| Parameters | Cubic $KB_2H_8$ | | | | | | |
|---|---|---|---|---|---|---|---|
| Pressure considered (GPa) | 0 | 4 | 8 | 12 | 16 | 20 | 24 |
| Energy tolerance (eV/atom) | $5.0\times10^{-6}$ | $5.0\times10^{-6}$ | $5.0\times10^{-6}$ | $5.0\times10^{-6}$ | $5.0\times10^{-6}$ | $5.0\times10^{-6}$ | $5.0\times10^{-6}$ |
| Max. force (eV/Å) | 0.01 | 0.01 | 0.01 | 0.01 | 0.01 | 0.01 | 0.01 |
| Max. stress (GPa) | 0.02 | 0.02 | 0.02 | 0.02 | 0.02 | 0.02 | 0.02 |
| Max. displacement (Å) | $5.0\times10^{-4}$ | $5.0\times10^{-4}$ | $5.0\times10^{-4}$ | $5.0\times10^{-4}$ | $5.0\times10^{-4}$ | $5.0\times10^{-4}$ | $5.0\times10^{-4}$ |
| Energy cut-off (eV) | 500 | 500 | 500 | 500 | 500 | 500 | 500 |
| $k$-points grid size | 16×16×16 | 16×16×16 | 16×16×16 | 16×16×16 | 16×16×16 | 16×16×16 | 16×16×16 |
| Valence electrons considered | H: $1s^1$ K: $3s^2\,3p^6\,4s^1$ B: $2s^2\,2p^1$ | | | | | | |

## 3. Results and analysis

### 3.1 Crystal structure and stability

Figure 1 shows the schematic crystal structure of cubic $KB_2H_8$. We have shown: (a) crystal structure in 2D view, (b) crystal structure in 3D view and (c) crystal structure of the Polyhedra. From these figures it is seen that each of B atom is surrounded by four H atoms and each $BH_4$ tetrahedron is surrounded by four potassium atoms, which also form a tetrahedron. The $KB_2H_8$ (space group *Fm-3m*) structure shown contains four formula units, i.e. there are 44 atoms in total in the unit cell. The optimized unit cell parameters (*a* and *V*), the cohesive energy $E_{coh}$, and the enthalpy of formation ($\Delta H$) at different hydrostatic pressures are summarized in Table 1.

In our work, we have calculated the cohesive energy per atom using the approach adopted in Refs.[33-35] to determine the thermodynamic stability. Furthermore, the cohesive $E_{coh}$ has been computed using the following equation:

$$E_{coh} = \frac{E_K + 2E_B + 8E_H - E_{KB_2H_8}}{11} \qquad (1)$$

where, $E_{KB_2H_8}$ is total energy per formula unit of $KB_2H_8$ and $E_K$, $E_B$ and $E_H$ are the total energies of single K, B and H atoms in the solid state, respectively. From Table 1, it is found that the values of cohesive energy per atom are positive and enthalpy of formation is negative, which indicate that the structure of $KB_2H_8$ is thermodynamically stable [36]. The structural parameters obtained here are in good agreement with the previous results [19].

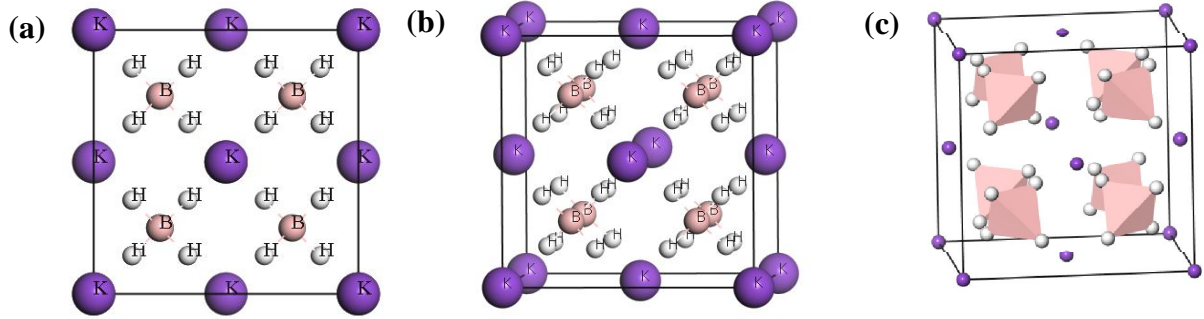

**Figure 1**: Crystal structure of cubic $KB_2H_8$: (a) 2D view, (b) 3D view and (c) Polyhedrons.

**Table 1**. Structural properties (*a* and *V*), cohesive energy/atom ($E_{coh}$), and enthalpy/atom ($\Delta H$) of cubic $KB_2H_8$.

| Pressure, $P$ (GPa) | Lattice parameters (Å) | | | Volume, $V$ (Å$^3$) | Cohesive energy, $E_{coh}$ (eV/atom) | Enthalpy, $\Delta H \times 10^2$ (eV/atom) | Ref. |
|---|---|---|---|---|---|---|---|
| | *a* | *b* | *c* | | | | |
| 0 | 6.96212 | 6.96212 | 6.96212 | 337.46 | 4.00 | -3.85055 | |
| 4 | 6.70041 | 6.70041 | 6.70041 | 300.82 | 3.99 | -3.84335 | |
| 8 | 6.53217 | 6.53217 | 6.53217 | 278.72 | 3.97 | -3.83678 | [This work] |
| 12 | 6.40427 | 6.40427 | 6.40427 | 262.67 | 3.95 | -3.83065 | |
| 16 | 6.30247 | 6.30247 | 6.30247 | 250.34 | 3.92 | -3.82483 | |
| 20 | 6.21513 | 6.21513 | 6.21513 | 240.08 | 3.90 | -3.81926 | |
| 24 | 6.14029 | 6.14029 | 6.14029 | 231.51 | 3.87 | -3.81391 | |

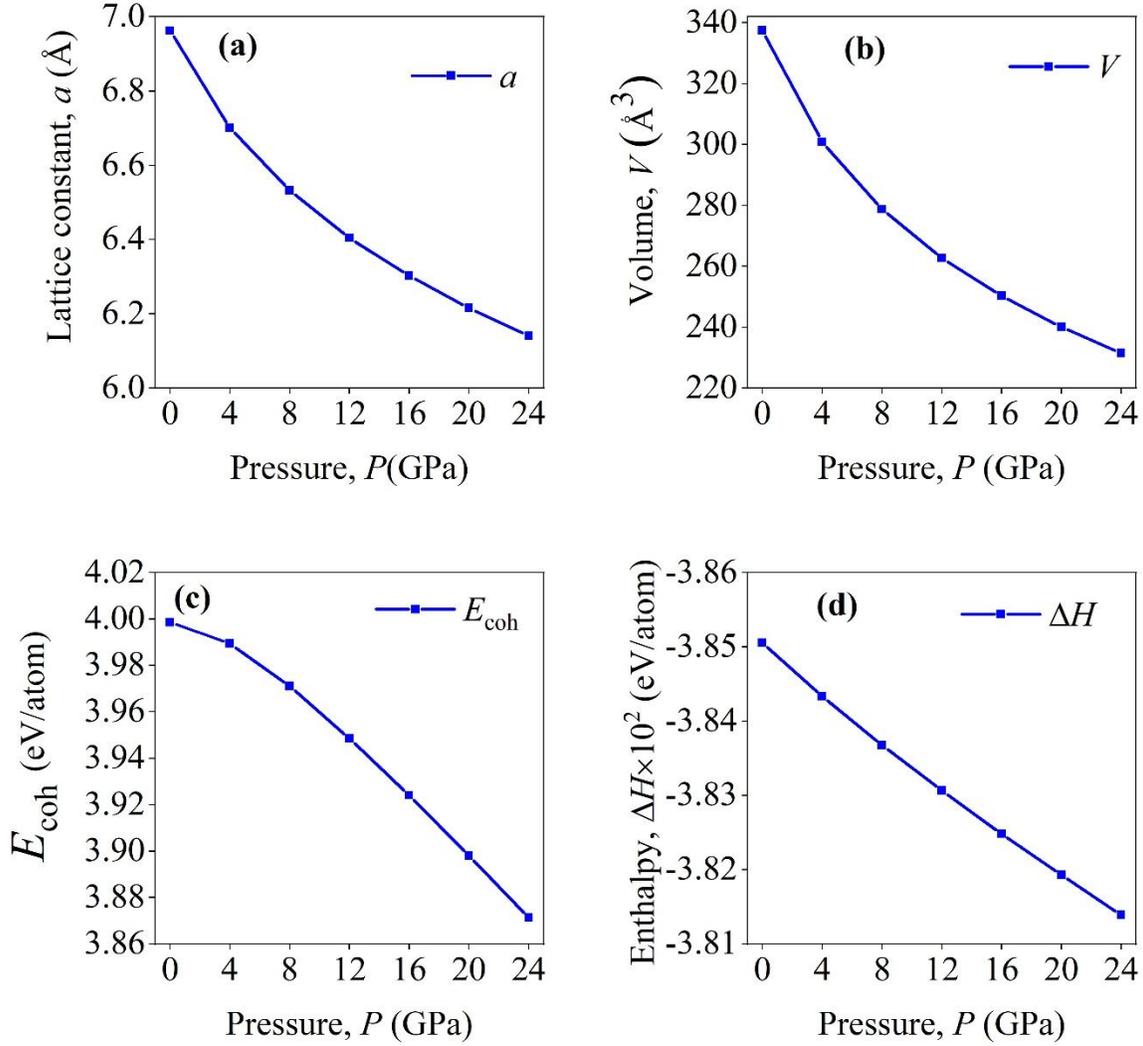

**Figure 2**: (a) Lattice parameters, (b) cell volume, (c) cohesive energy/atom and (d) enthalpy/atom of cubic $KB_2H_8$ under different pressures.

From Table 1 (cf. Fig. 1) it is observed that structural parameters ($a = b = c$, and $V$) decrease with pressure. The cohesive energy and the enthalpy also decrease with increasing pressure. We don't see any abrupt change in structural parameters or in cohesive energy/enthalpy in the pressure range considered. This suggests that there is no signature of thermodynamic instability in this pressure range.

## 3.2 Elastic properties

### 3.2.1 Single crystal elastic constants

For a cubic crystal, there are three independent elastic constants: $C_{11}$, $C_{12}$, and $C_{44}$. All of these elastic constants are computed and are listed in Table 2 for the $KB_2H_8$ compound under different pressures. From Table 2 it is seen that $C_{ij}$ increases almost linearly with pressure in the range 8 GPa to 24 GPa. From Table 2 (cf. Fig. 3 (a)) it is also observed that elastic constant $C_{ij}$ are anomalous for pressures up to 4 GPa. This suggests that the cubic $KB_2H_8$ compound is elastically unstable in this pressure range. All the single crystal elastic constants $C_{ij}$ are positive for pressures 8 to 24 GPa. The values of $C_{11}$, $C_{22}$, and $C_{33}$ represent the directional resistance against uniaxial stress. The cubic symmetry of the compound demands $C_{11} = C_{22} = C_{33}$. The stiffness constants $C_{12}$ and $C_{44}$ represent the resistance against shearing strain for shears along different crystal planes. All these stiffness constants are determined by the interatomic bonding strength and the arrangement of the atoms in the unit cell. In the presence of hydrostatic pressure, the elastic stability conditions of a cubic crystal are: $(C_{11} - P) > 0$; $(C_{11} - P) > |(C_{12} + P)|$; $(C_{11} + 2C_{12} + P) > 0$ [37]. From Table 2 it is observed that the requirements of stability criteria under pressure are satisfied for pressures of 4 GPa to 24 GPa. The facts that, $C_{11} > C_{12}$, and $C_{11} > C_{44}$, indicate axial bonding between nearest neighbor atoms is stronger than the non-axial bondings. On the other hand, the Cauchy pressure (CP) given by $(C_{12} - C_{44})$, defines the ductility of a materials, because it is related to the angular character of atomic bonding in a solid [38]. In general, CP > 0 indicates ductile nature and CP < 0 indicates brittle nature. From Table 2, it is observed that cubic $KB_2H_8$ is a ductile solid in the pressure range considered in our calculations. The value of the tetragonal shear modulus (TSM) given by $(C_{11} - C_{12})/2$, is used to measure the shear stiffness of a crystal. TSM is also linked with the sound velocity in the solid. Positive values of TSM are indicative of dynamical stability of the structure in the long wavelength limit. The values of TSM indicate dynamical stability of $KB_2H_8$ in the pressure range 8 - 24 GPa.

**Table 2.** Single crystal elastic constants ($C_{ij}$), Cauchy pressure (CP) ($C'$) and tetragonal shear modulus (TSM) ($C''$) of cubic $KB_2H_8$.

| Pressure, $P$ (GPa) | $C_{ii}$ (GPa) | | | CP (GPa) | TSM (GPa) | Ref. |
|---|---|---|---|---|---|---|
| | $C_{11}$ | $C_{12}$ | $C_{44}$ | $C'$ | $C''$ | |
| 0  | -192.76 | 145.84 | 12.36 | 133.48 | -169.30 | |
| 4  | 26.15   | 53.35  | 17.81 | 35.54  | -13.60  | |
| 8  | 65.33   | 58.32  | 20.83 | 37.49  | 3.50    | |
| 12 | 97.04   | 64.89  | 24.65 | 40.24  | 16.08   | [This work] |
| 16 | 123.75  | 72.12  | 25.79 | 46.34  | 25.82   | |
| 20 | 150.23  | 80.34  | 26.81 | 53.54  | 34.94   | |
| 24 | 173.64  | 88.98  | 29.67 | 59.31  | 42.33   | |

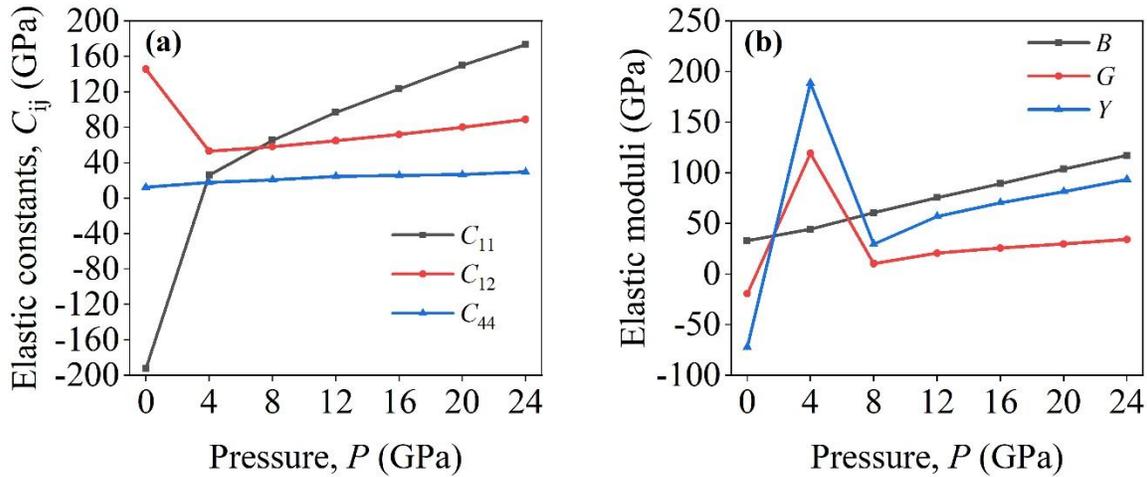

**Figure 3**: (a) Single crystal elastic constants and (b) Elastic moduli under pressure.

*3.2.2 Polycrystalline elastic properties*

The values of polycrystalline elastic moduli (bulk modulus, $B$, shear modulus $G$, and Young's modulus $Y$) are tabulated in Table 3. These elastic moduli were obtained from the computed values of $C_{ij}$ using the Voigt-Reuss-Hill (VRH) approximation [39-41]. Hill approximated elastic moduli are the average of the Voigt and Reuss approximated elastic moduli [41] which represents the real situation quite accurately. From Table 3 it is observed that Bulk modulus increases systematically from 0-24 GPa but the shear and Young's modulus increases systematically only from 8-24 GPa. The bulk modulus is the indicator of the elastic resistance against uniform volume distortion. Large value of $B$ implies high incompressibility. Bulk modulus is also an average measure of the strength of interatomic bondings. The shear modulus defines the elastic resistance against shape changing strains. High shear modulus is indicative of strong directional chemical bonding. From Table 3, it is observed that the resistance to shape change increases with pressure of from 8 GPa to 24 GPa. Young's modulus represents the resistance against tensile strain. Negative values of $G$ and $Y$ for cubic $KB_2H_8$ at zero pressure suggest again the elastic instability. The anomalously high value of Young's modulus and modulus of rigidity at 4 GPa also implies mechanical instability. It is worth mentioning that Gao et al. [19] also found $KB_2H_8$ to be mechanically and dynamically stable only at high pressures.

**Table 3**. Polycrystalline elastic parameters of cubic $KB_2H_8$ under different pressures.

| Pressure, $P$ (GPa) | Polycrystalline elastic moduli (GPa) | | | | | | | | | Ref. |
|---|---|---|---|---|---|---|---|---|---|---|
| | Bulk, $B$ | | | Shear, $G$ | | | Young's, $Y$ | | | |
| | $B_V$ | $B_R$ | $B_H$ | $G_V$ | $G_R$ | $G_H$ | $Y_V$ | $Y_R$ | $Y_H$ | |
| 0 | 32.98 | 32.98 | 32.98 | -60.30 | 21.66 | -19.32 | -463.31 | 53.31 | -72.03 | |
| 4 | 44.29 | 44.29 | 44.29 | 5.25 | 233.56 | 119.40 | 15.14 | 254.06 | 188.66 | |
| 8 | 60.66 | 60.66 | 60.66 | 13.90 | 7.00 | 10.45 | 38.74 | 20.21 | 29.64 | |
| 12 | 75.61 | 75.61 | 75.61 | 21.22 | 20.32 | 20.77 | 58.22 | 55.94 | 57.08 | [This work] |
| 16 | 89.33 | 89.33 | 89.33 | 25.80 | 25.80 | 25.80 | 70.60 | 70.60 | 70.60 | |
| 20 | 103.64 | 103.64 | 103.64 | 30.06 | 29.56 | 29.81 | 82.23 | 80.98 | 81.61 | |
| 24 | 117.20 | 117.20 | 117.20 | 34.73 | 33.70 | 34.22 | 94.83 | 92.26 | 93.55 | |

**Table 4**. Calculated Poisson's ratio ($v$), Pugh's ratio ($G/B$) and machinability index ($\mu_m$) of cubic $KB_2H_8$ at different pressures.

| Pressure, $P$ (GPa) | $v$ | $G/B$ | $\mu_m$ | Ref. |
|---|---|---|---|---|
| 0 | 0.86 | -0.59 | 2.67 | |
| 4 | -0.21 | 2.70 | 2.49 | |
| 8 | 0.42 | 0.23 | 2.91 | |
| 12 | 0.37 | 0.27 | 3.07 | [This work] |
| 16 | 0.37 | 0.29 | 3.46 | |
| 20 | 0.37 | 0.29 | 3.87 | |
| 24 | 0.37 | 0.29 | 3.95 | |

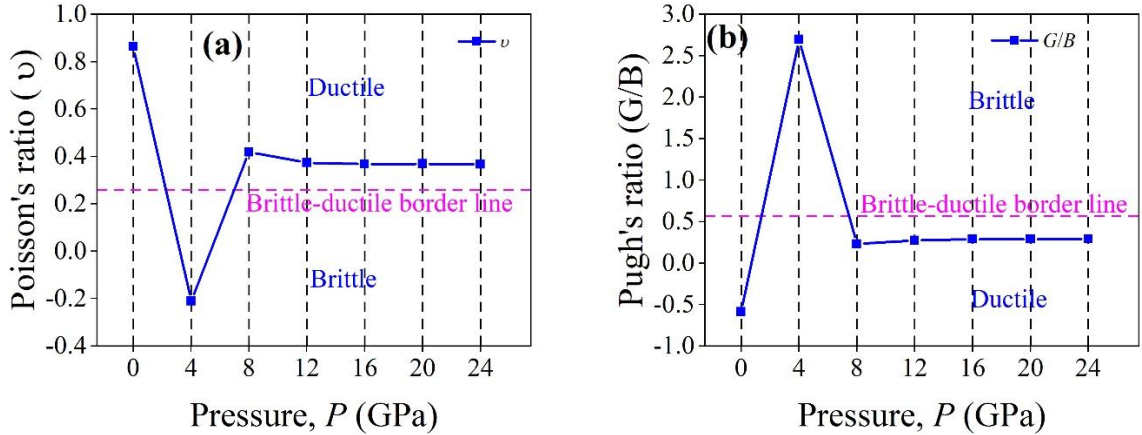

**Figure 4**: (a) Poisson's ratio and (b) Pugh's ratio of $KB_2H_8$ under pressure.

The Poisson's and Pugh's ratios [38] define the brittle and ductile nature of a compound. If the Poisson's ratio, $v$, of a solid is greater than 0.26, it is expected to be ductile and, on the other hand, if $v < 0.26$, it should exhibit brittleness. From Table 4 it is observed that the Poisson's ratio of $KB_2H_8$ varies widely in the pressure range from 0 GPa to 4 GPa. Above 4 GPa, a systematic behavior is found. The high values of Poisson's ratio imply ductility and presence of significant non-directional ionic and metallic bondings. The same qualitative trend is seen in the Pugh's ratio, $G/B$. For the Pugh's ratio, if $G/B > 0.57$, the solid exhibits brittleness; otherwise it is ductile. From Table 4 (cf. Fig. 4), we find that cubic $KB_2H_8$ is ductile in the pressure range 8 – 24 GPa. The machinability index, $\mu_m$, is an important mechanical parameter giving information about the ease with which a solid can be shaped and cut into desired geometry. It is also used as an indicator of dry lubricity of solids. High machinability index indicates the ease in cutting and low level of frictional loss. The machinability index increases systematically with increasing pressure from 8 GPa to 24 GPa. From Table 4 it is observed that at high pressures cubic $KB_2H_8$ has very high level of machinability and excellent dry lubricity [42].

*3.3 Elastic Anisotropy*

Information on elastic anisotropy is essential for engineering applications of solids. The elastic anisotropy determines the direction dependent mechanical properties of solids under external stress. The elastic anisotropy reflects the anisotropy in the bonding strengths between atoms located in different directions and different planes.

**Table 5**. The shear anisotropy factors $A_1$, $A_2$, $A_3$, and $A^B$ (in %), $A^G$ (in %), $A^U$ of cubic $KB_2H_8$ under different pressures.

| Pressure, $P$ (GPa) | $A_1$ | $A_2$ | $A_3$ | $A^B$ | $A^G$ | $A^U$ | Ref. |
|---|---|---|---|---|---|---|---|
| 0 | -0.07 | -0.07 | -0.07 | 0.00 | 212.10 | -18.92 | |
| 4 | -1.31 | -1.31 | -1.31 | 0.00 | -95.61 | -4.89 | |
| 8 | 5.94 | 5.94 | 5.94 | 0.00 | 33.03 | 4.93 | |
| 12 | 1.53 | 1.53 | 1.53 | 0.00 | 2.18 | 0.22 | [This work] |
| 16 | 1.00 | 1.00 | 1.00 | 0.00 | 0.00 | 0.00 | |
| 20 | 0.77 | 0.77 | 0.77 | 0.00 | 0.84 | 0.08 | |
| 24 | 0.70 | 0.70 | 0.70 | 0.00 | 1.51 | 0.15 | |

The shear anisotropy factors considering different crystal planes, $A_1$, $A_2$, and $A_3$ are all equal to one for an isotropic crystal, while any value different from unity is a measure of the degree of shear elastic anisotropy possessed by the crystal. The factors $A_1$, $A_2$, and $A_3$ are computed using the following equations [43,44]:

$$A_1 = \frac{4C_{44}}{C_{11} + C_{33} - 2C_{13}}, A_2 = \frac{4C_{55}}{C_{22} + C_{33} - 2C_{23}}, A_3 = \frac{4C_{66}}{C_{11} + C_{22} - 2C_{12}} \quad (2)$$

Again, the anisotropy factors $A^B$ and $A^G$ are the percentage anisotropies in compressibility and shear, respectively. $A^U$ is the universal anisotropy index. The zero values of $A^B$, $A^G$ and $A^U$ represent elastic isotropy of a crystal and non-zero values of these indicators represent anisotropy [43,44]. The factors $A^B$, $A^G$ and $A^U$ are calculated using the following equations:

$$A^B = \frac{B_V - B_R}{B_V + B_R}, A^G = \frac{G_V - G_R}{G_V + G_R}, \text{ and } A^U = 5\frac{G_V}{G_R} + \frac{B_V}{B_R} - 6 \geq 0 \quad (3)$$

The computed values of $A_1$, $A_2$, $A_3$ $A^U$, $A^B$, and $A^G$ are listed in Table 5. Table 5 shows clearly that the cubic $KB_2H_8$ is elastically anisotropic. The pressure dependent variations in various anisotropy indices are nonmonotonic, indicating that changes in the bonding characters are different in different crystal directions when uniform hydrostatic pressure is applied. The negative values of $A^U$ for pressures 0 GPa and 4 GPa once again imply that $KB_2H_8$ is structurally unstable at low pressures. In the subsequent tables, we have not shown any data for 0 GPa and 4 GPa, since at these pressures the structure of $KB_2H_8$ gives unphysical results. Similar conclusions have been drawn in previous studies [19].

## 3.4 Sound velocities and hardness

A large number of thermophysical parameters of a crystal is closely related with the sound velocities [45]. Crystalline solids support both longitudinal and transverse modes of propagation of acoustic disturbances. The phonon thermal conductivity of solids and sound velocities increases following the same trend. The calculated values of sound velocities ($v_t$, $v_l$ and $v_m$) under pressure of $KB_2H_8$ are listed in Table 6.

**Table 6.** Calculated density ($\rho$), transverse sound velocity ($v_t$), longitudinal sound velocity ($v_l$) and average sound velocity ($v_m$) of $KB_2H_8$.

| Pressure, $P$ (GPa) | Density, $\rho$ (gm/cm³) | Sound velocities (km/s) | | | Ref. |
|---|---|---|---|---|---|
| | | $v_t$ | $v_l$ | $v_m$ | |
| 8  | 1.639 | 2.9119 | 6.9506 | 3.2934 | |
| 12 | 1.739 | 3.4556 | 7.7067 | 3.8980 | |
| 16 | 1.825 | 3.7599 | 8.2341 | 4.2378 | [This work] |
| 20 | 1.903 | 3.9580 | 8.6804 | 4.4614 | |
| 24 | 1.973 | 4.1640 | 9.0834 | 4.6925 | |

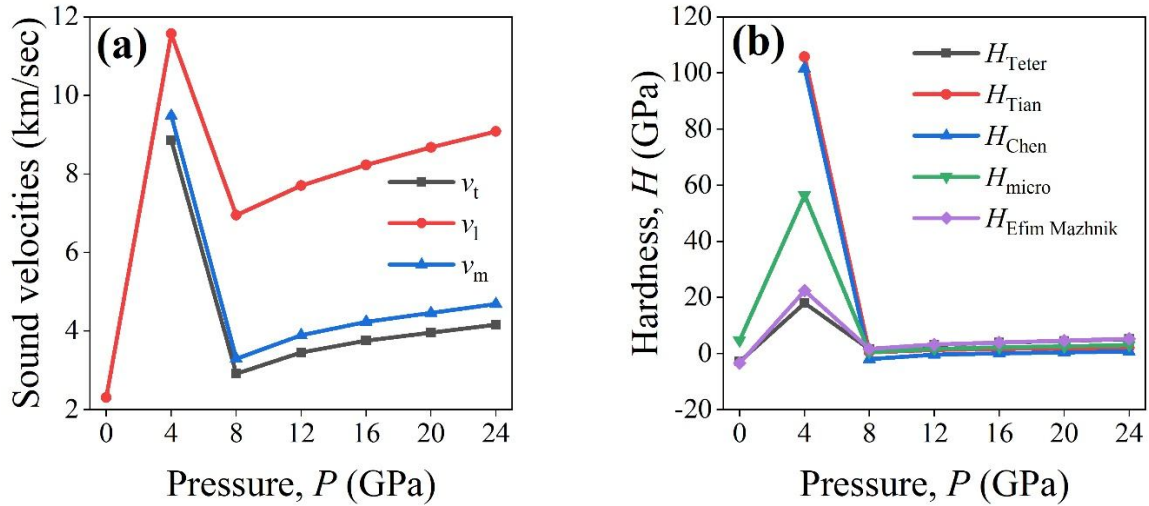

**Figure 5**: (a) Sound velocities and (b) Hardness of $KB_2H_8$ under pressure.

The sound velocities were determined using the following equations [46-48]:

$$v_t = \sqrt{\frac{G}{\rho}}, \; v_l = \sqrt{\frac{3B + 4G}{3\rho}} \text{ and } v_m = \left[\frac{1}{3}\left(\frac{2}{v_t^3} + \frac{1}{v_l^3}\right)\right]^{-\frac{1}{3}} \tag{4}$$

where, $v_t$, $v_l$ and $v_m$ signify the transverse, longitudinal, and the mean sound velocities, respectively. The computed values of sound velocities under pressures are given in Table 6 and are also illustrated in Fig. 5 (a). In general, the sound velocities increase with increasing pressure due to the increase in the crystal stiffness and density [49]. This is seen from 8 GPa to 24 GP. The longitudinal sound velocities are larger than that of transverse velocity for each direction of propagation. This is due to the fact that $C_{11}$ is greater than $C_{12}$ or $C_{44}$ for $KB_2H_8$. The directional sound velocities [49] are also calculated. The direction dependent acoustic velocities along different crystallographic axes are determined by the atomic arrangements and bonding strengths in the specific directions. In Table 7, we have displayed the data in the pressure range 8 – 24 GPa, since at lower pressures the structure is unstable.

**Table 7**. The longitudinal $v_l$ (in ms$^{-1}$) and transverse wave velocities ($v_{t_1}$ and $v_{t_2}$ in ms$^{-1}$) along [100], [110] and [111] directions in cubic $KB_2H_8$ under pressure.

| Pressure, $P$ (GPa) | [100] | | [110] | | | [111] | |
|---|---|---|---|---|---|---|---|
| | $v_l$ | $v_t$ | $v_l$ | $v_{t_1}$ | $v_{t_2}$ | $v_l$ | $v_t$ |
| 8 | 650.256 | 3564.64 | 7101.01 | 3564.64 | 1462.28 | 7344.87 | 2379.3 |
| 12 | 769.371 | 3764.83 | 7792.73 | 3764.83 | 3040.16 | 7897.5 | 3299.45 |
| 16 | 848.192 | 3758.97 | 8233.86 | 3758.97 | 3761.19 | 8233.52 | 3760.45 |
| 20 | 915.153 | 3753.45 | 8641.19 | 3753.45 | 4284.98 | 8558.38 | 4115.43 |
| 24 | 966.181 | 3877.33 | 9031.84 | 3877.33 | 4631.57 | 8912.63 | 4394.57 |

Mechanical hardness is used to evaluate the elastic and plastic behavior of a crystal. In our study, we have calculated the hardness of the $KB_2H_8$ at different pressures using the formalisms developed by Teter et al. [50], Tian et al. [51], Chen et al. [52], and microhardness [53], and the one developed by Efim Mazhnik [54]. The calculated values of hardness are given in Table 8. Within the formalisms of Teter et al. [50] and Chen et al. [52], hardness depends on shear and bulk modulus. We also know that higher hardness of a solid suggests higher Debye temperature. Our calculated results agree with this (Tables 8 and 9). Hardness is also closely related to sound velocity. From Table 8 (cf. Fig. 5) it observed that hardness and sound velocities are increasing following the same trend with pressure.

**Table 8.** Hardness of cubic $KB_2H_8$ under pressure.

| Pressure, $P$ (GPa) | Hardness $H$ (GPa) | | | | | Ref. |
|---|---|---|---|---|---|---|
| | $H_{Teter}$ | $H_{Tian}$ | $H_{Chen}$ | $H_{micro}$ | $H_{Efim\ Mazhnik}$ | |
| 8 | 1.58 | 0.44 | -1.99 | 0.57 | 1.71 | |
| 12 | 3.14 | 1.34 | -0.40 | 1.74 | 3.26 | |
| 16 | 3.90 | 1.68 | 0.13 | 2.27 | 4.02 | [This work] |
| 20 | 4.50 | 1.85 | 0.39 | 2.61 | 4.65 | |
| 24 | 5.17 | 2.08 | 0.74 | 3.03 | 5.33 | |

It is worth noting that, both hardness and superconductivity are closely related to the electronic band structure and the strength of the chemical bonds. Therefore, a change in the hardness with pressure of a solid may have an effect on the superconducting state properties, including the critical temperature.

*3.5 Thermophysical properties*

We have calculated various thermophysical parameters in this section. The computed results are tabulated below (Table 9).

**Table 9.** Calculated Grüneisen parameter ($\gamma$), Kleinman parameter ($\zeta$), Debye temperature $\theta_D$ (K) and melting temperature $T_m$ (K) of $KB_2H_8$.

| Pressure, $P$ (GPa) | $\gamma$ | $\zeta$ | $\theta_D$ | $T_m$ | Ref. |
|---|---|---|---|---|---|
| 8 | 2.86 | 0.93 | 529.97 | 647.98 | |
| 12 | 2.35 | 0.76 | 639.77 | 790.70 | |
| 16 | 2.29 | 0.69 | 706.77 | 910.90 | [This work] |
| 20 | 2.30 | 0.65 | 754.53 | 1030.02 | |
| 24 | 2.28 | 0.64 | 803.28 | 1135.39 | |

*Grüneisen parameter* ($\gamma$): The Grüneisen parameter, $\gamma$ is an important parameter which is closely related with the vibrational properties of solids. It is related also to the thermal expansion coefficient, bulk modulus, specific heat, and electron-phonon coupling in solids. The normal thermal expansion of solids due to anharmonicity of interatomic forces is understood from the Grüneisen constant as well. Both the Grüneisen parameter and the Poisson ratio $v$ (the lateral strain coefficient) characterize the tendency of a material towards retaining its initial volume in the course of elastic deformation and are closely related. The relation between Grüneisen parameter and Poisson's ratio is as follows: $\gamma = \frac{3}{2}\frac{1+v}{2-3v}$ [55]. The lower limit of Poisson's ratio, i.e $v = -1.0$ correspond to a completely harmonic solid. High value of Grüneisen parameter is indicative of strong anharmonicity in lattice dynamics and can lead to significant electron-phonon interaction in superconductors [56]. The pressure dependent values of the Grüneisen parameter of $KB_2H_8$, calculated using the Poisson's ratio, are given in Table 9. The obtained values of $\gamma$ for cubic $KB_2H_8$ are high [57]. The calculated values of the Grüneisen parameter are decreasing with increasing pressure (Table 9).

*Kleinman parameter* ($\zeta$): This parameter was introduced by Kleinman [58] describing the relative ease of bond bending versus the bond stretching. Minimum contribution from bond bending leads to $\zeta = 0$ and minimum contribution from bond stretching leads to $\zeta = 1.0$. The Kleinman parameter is calculated using the elastic constants $C_{11}$ and $C_{12}$ as follows:

$$\zeta = \frac{C_{11} + 8C_{12}}{7C_{11} + 2C_{12}} \tag{5}$$

The computed values of $\zeta$ are displayed in Table 9. It is clear from the table that the bond bending contributions dominate in the elastic properties of cubic $KB_2H_8$. The values of $\zeta$ decrease systematically with increasing pressure.

*Debye temperature* ($\theta_D$): Debye temperature $\theta_D$ is closely related to many physical properties of solids such as specific heat, melting temperature, thermal conductivity, hardness of solids, elastic constants, acoustic velocity etc. Moreover, Debye temperature provides information about the electron-phonon coupling and Cooper pairing mechanism in superconductivity. At low temperatures, the vibrational excitations arise solely from acoustic vibrations. Thus the Debye temperature calculated from the elastic constants is considered to be similar as that acquired from the specific heat measurements. In this work, the Debye temperature has been calculated using the Anderson method [59] as follows:

$$\theta_D = \frac{h}{k_B}\left[\left(\frac{3n}{4\pi}\right)\frac{N_A\rho}{M}\right]^{1/3} v_m \tag{6}$$

where, $h$ is Planck's constant, $k_B$ is the Boltzmann's constant, $\rho$ is the density, $N_A$ is the Avogadro number, $M$ is the molecular mass, and $v_m$ is the average sound velocity. The mean sound velocity can be determined from the Eqn. 4.

From Table 9, it can be inferred that the $\theta_D$ increases monotonously from 8 GPa. Usually, the increase of $\theta_D$ with pressure indicates the crystal stiffening, but in the opposite case the system is driven effectively towards lattice softening. The anomalous behavior of $G$ under pressure may be responsible for the anomalous change in $\theta_D$ (cf. Table 3 and Table 9). At this point, we would like to stress that the high predicted $T_C$ of $KB_2H_8$ is partly due to its reasonably high Debye temperature. In conventional phonon mediated superconductors, the superconducting critical temperature is directly proportional to the Debye temperature. The pressure induced increase in the Debye temperature found for $KB_2H_8$ suggests that $T_C$ should increase markedly with increasing pressure provided that the electron-phonon coupling constant does not decrease significantly with pressure variation.

*Melting temperature* ($T_m$): The melting temperature of cubic compound is calculated with an empirical formula based on single crystal elastic constants [60]:

$$T_m = 354 + 1.5(2C_{11} + C_{33}) \tag{7}$$

The calculated $T_m$ is listed in Table 9. It is found that the melting point increases almost linearly with the increase of pressure. Compound with high melting temperature has lower thermal expansion and high overall bonding energy (cf. Table 9 and Table 1) and also has high Debye temperature (cf. Table 9). The melting temperatures of $KB_2H_8$ are quite low, consistent with the relatively soft nature of the compound (Table 8).

*Minimum phonon thermal conductivity and thermal expansion coefficient*: Thermal conductivity is a measure of the ability of a system for conducting heat energy provided that there is a temperature gradient. Thermal conductivity is an important topic in that the magnitude and temperature dependence of this parameter yield valuable information about the material under study. Specifically, it sheds light on the electronic and vibrational states of the material, interactions between the different heat conducting entities within the system, and on the structural integrity of the material through which the thermal energy propagates. In general, in metallic systems, heat is transported via electrons and phonons. At high temperatures the phonon contribution dominates. In this section we have presented the results of calculations of the minimum phonon thermal conductivity, and the thermal expansion coefficient (TEC) of cubic $KB_2H_8$ for pressures in the range 8 – 24 GPa. The calculated values are given in Table 10 below. We have also presented the pressure dependent values of fracture toughness of the compound under study in this table.

**Table 10**. Thermal expansion coefficient $\alpha$ ($10^{-5}$ K$^{-1}$), minimum thermal conductivity, $k_{min}$ (Wm$^{-1}$K$^{-1}$), and the fracture toughness, $K_{IC}$ (MPam$^{-1/2}$) of cubic $KB_2H_8$.

| Pressure, $P$ (GPa) | $\alpha$ | $k_{min}$ | | $K_{IC}$ | | Ref. |
|---|---|---|---|---|---|---|
| | | Cahill | Clark | Mazhnik | Kvashnin | |
| 8 | 11.51 | 2.08 | 1.33 | 0.18 | 0.34 | [This work] |
| 12 | 7.70 | 2.47 | 1.64 | 0.33 | 0.53 | |
| 16 | 6.20 | 2.75 | 1.84 | 0.44 | 0.64 | |
| 20 | 5.37 | 2.98 | 1.99 | 0.54 | 0.74 | |
| 24 | 4.68 | 3.20 | 2.14 | 0.65 | 0.84 | |

The TEC of a material is interconnected to many other physical properties, such as thermal conductivity, specific heat and temperature variation of the semiconducting energy band gap of materials. This parameter ($\alpha$) is also important for growth of crystals. The TEC of a solid is inversely proportional to the modulus of rigidity and can be estimated using the following relation [61]:

$$\alpha = \frac{1.6 \times 10^{-3}}{G} \tag{8}$$

The approximate relation between thermal expansion coefficient and melting temperature is as follows $\alpha \approx 0.02/T_m$ [61] i.e., thermal expansion coefficient is inversely proportional to the melting temperature. This is a natural consequence of the dependences of these two parameters on the average interatomic bonding strength. Our calculated values are in good agreement with this approximate relation.

At temperatures above Debye temperature, the phonon thermal conductivity approaches a minimum value, known as minimum thermal conductivity. It is denoted by $k_{min}$. According to

Cahill and Clarke model minimum thermal conductivity can be calculated using the following equation [62]:

$$k_{min} = \frac{k_B}{2.48} n^{\frac{2}{3}}(v_l + v_{t1} + v_{t2}) \qquad (9)$$

It is clear that the minimum thermal conductivity depends on different sound velocities in different crystallographic directions. From our calculations it is observed that sound velocities increase (cf. Table 6 and Table 7) as well as the minimum thermal conductivity increases under pressure. Clark further deduced the following equation [63] to calculate the bulk minimum thermal conductivity:

$$k_{min} = k_B v_a (V_{atomic})^{-\frac{2}{3}} \qquad (10)$$

In these equations $k_B$ is the Boltzmann constant, $v_a$ is the average sound velocity and $V_{atomic}$ is the cell volume per atom.

For industrial use of materials, fracture toughness as well as hardness plays important roles. The fracture toughness describes the resistance of a solid to prevent the propagation of a crack produced inside. Quantitatively, it can be determined from the stress intensity factor $K$ at which a thin crack in the material begins to spread. According to Efim Mazhnik et al. [54,64] facture toughness can be calculated using the following equation:

$$K_{IC} = \alpha_0^{-\frac{1}{2}} V_0^{\frac{1}{6}} \xi(v) Y^{\frac{3}{2}} \qquad (11)$$

where, $\alpha$ = 8840, $V_0$ is the volume per atom, $\xi(v)$ is a dimensionless function of Poisson's ratio defined as follows:

$$\xi(v) = \frac{1 - 13.7v + 48.6v^2}{1 - 15.2v + 70.2v^2 - 81.5v^3} \qquad (12)$$

Besides, Alexander G. Kvashnin et al. [65] used the following equation for calculating the fracture toughness:

$$K_{IC} = \alpha V_0^{\frac{1}{6}} G \left(\frac{B}{G}\right)^{\frac{1}{2}} \qquad (13)$$

where, $\alpha = 1$ for the cubic hydride material, $V_0$ is the volume per atom, $G$ and $B$ are the shear and bulk moduli. A material should have high fracture toughness along with high hardness for heavy duty industrial applications; consequently prediction of the fracture toughness has drawn significant engineering interest. From Table 10, we see that the fracture toughness increases significantly with increasing pressure.

*3.6 Electronic band structure and properties*

The calculated electronic band structures at equilibrium lattice constants with different pressures are drawn along high symmetry directions in the first Brillouin zone and are illustrated in Figs. 6. We can see from these figures that there is no band gap and three distinct bands cross the Fermi level. This indicates the metallic nature of cubic $KB_2H_8$ for the pressures considered. The gross features of the band structures agree quite well with previous findings [19].

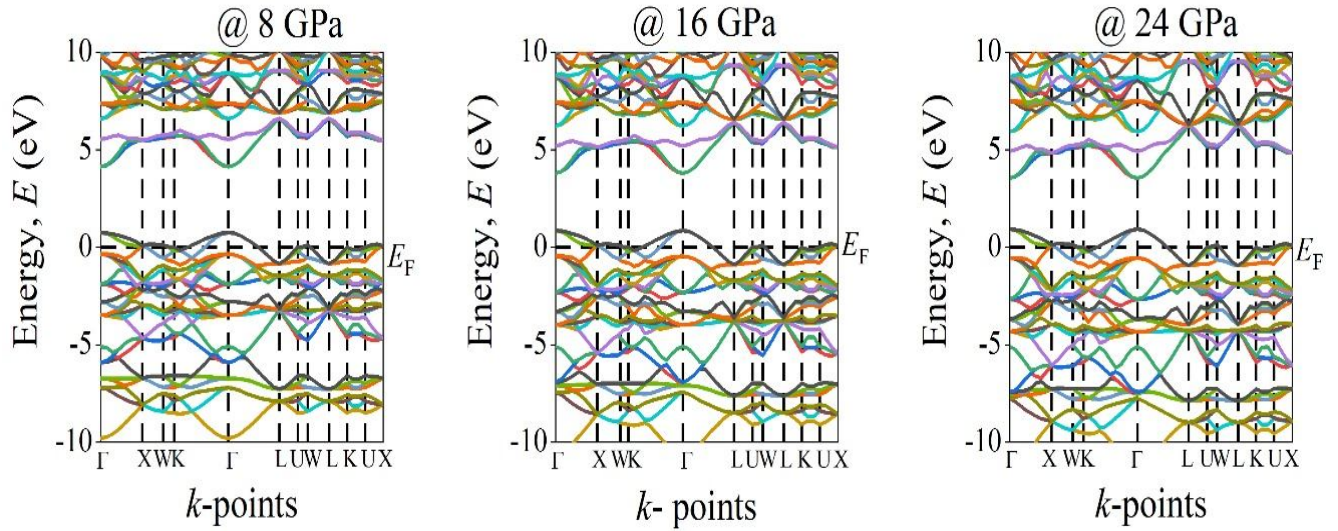

**Figure 6**: Electronic band structure of $KB_2H_8$ under different pressures.

The degree of electronic dispersion is direction dependent within the Brillouin zone. From Fig. 6 it is seen that the bands crossing the Fermi level become more dispersive as pressure increases. Thus, the charge carrier effective mass should decrease gradually with increasing pressure.

The total and partial density of states (TDOS and PDOS, respectively) are shown in Fig. 7. The values of TDOS at the Fermi level are 5.38, 4.93, 4.68, 4.50 and 4.40 states/eV-unit cell at 8 GPa, 12 GPa, 16 GPa, 20 GPa and 24 GPa, respectively. The values of non-zero TDOS at the Fermi energy again confirm the metallic nature of $KB_2H_8$. The TDOS close to the Fermi level originated mainly from H-$s$ states and B-$p$ states. Similar results were found in a previous study [19]. The contribution of the electronic orbitals of the K atom to the TDOS is very small.

The values of DOS at Fermi level $N(E_F)$ (states/eV-formula unit) is an extremely important electronic parameter that controls a large number of charge transport and magnetic properties of metals. Besides, the repulsive Coulomb pseudopotential, $\mu^*$ is also determined by the TDOS at the Fermi level. The computed values of $\mu^*$ at different pressures are shown in Table 11. We have calculated $\mu^*$ using the following equation [66,67]:

$$\mu^* = \frac{0.26 N(E_F)}{1 + N(E_F)} \tag{14}$$

In this equation, the TDOS has to be expressed considering the formula unit. It is observed from Table 11 that $\mu^*$ decreases slowly with rising pressure. This implies that the Coulomb electronic correlations become weaker with increasing pressure. The repulsive Coulomb pseudopotential has significant effect on superconducting transition temperature. This parameter weakens the attractive electron-phonon interaction, essential for the formation of Cooper pairs, and reduces the superconducting critical temperature [67,68].

**Table 11**. TDOS at the Fermi level $N(E_F)$ (states/eV-formula unit) and repulsive Coulomb pseudopotential $\mu^*$ of $KB_2H_8$ under different pressures.

| Pressure, P (GPa) | $N(E_F)$ | $\mu^*$ | Ref. |
|---|---|---|---|
| 8 | 1.39 | 0.151 | |
| 12 | 1.23 | 0.143 | |
| 16 | 1.17 | 0.140 | [This work] |
| 20 | 1.12 | 0.137 | |
| 24 | 1.10 | 0.136 | |

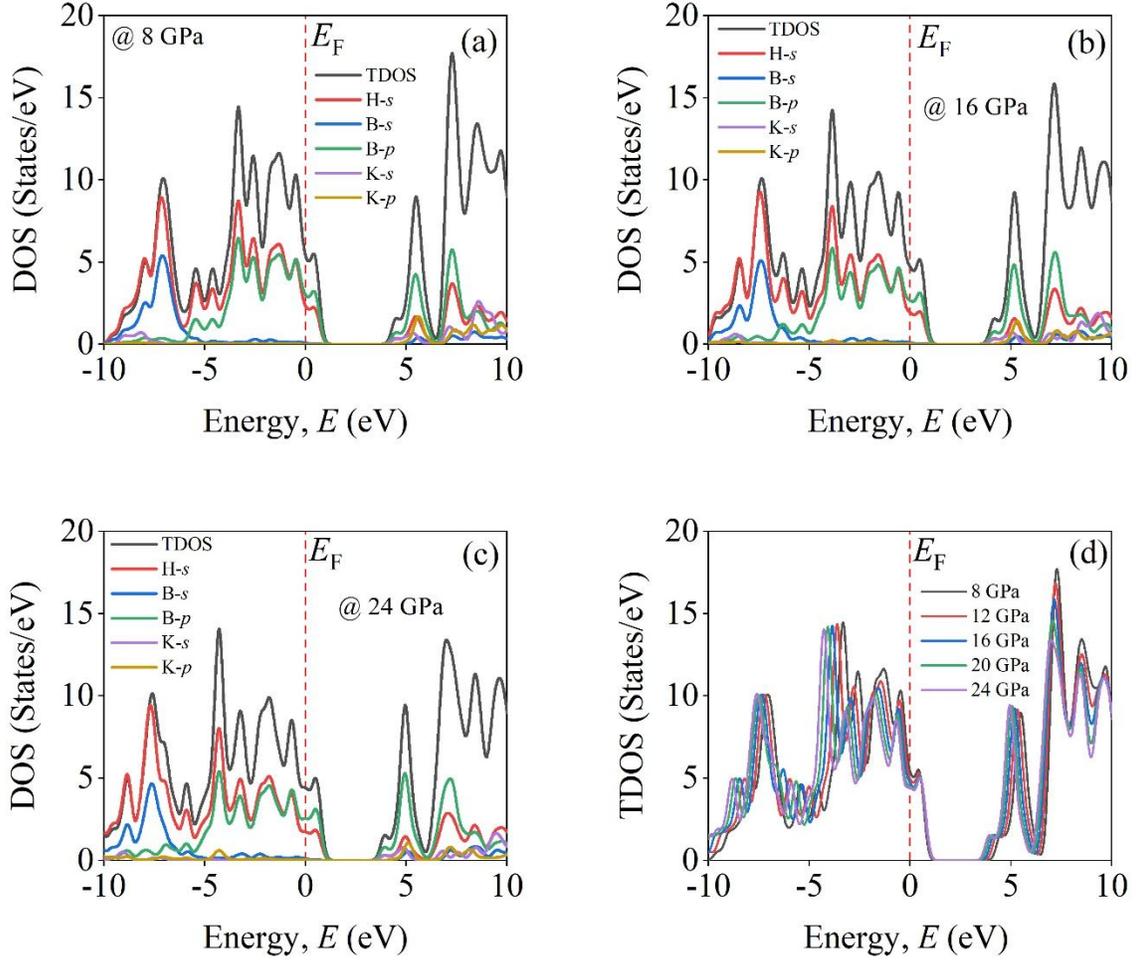

**Figure 7**: Density of states (TDOS and PDOS) of cubic $KB_2H_8$ under different pressures. Panel (d) summarizes all the plots together.

It is interesting to note from Fig. 7d that the peaks in the TDOS are shifted systematically to lower energies with increasing pressure. There is also significant hybridization between the H-$s$ and B-$p$ electronic states just below the Fermi energy. These states are primarily involved in the atomic bondings within $KB_2H_8$.

*3.7 Superconducting state properties*

In a previous work, Gao et al. [19] predicted high-$T_C$ superconductivity in cubic $KB_2H_8$ under pressure. Taking the repulsive Coulomb pseudopotential within 0.10 to 0.15, Gao et al. [19] calculated the superconducting critical temperature within the range from 134 K to 146 K by solving the self-consistent Eliashberg equation. In the same work, it was found that the electron-phonon coupling constant, $\lambda_{ep}$ decreases systematically with increasing pressure. To be specific,

Gao et al. [19] found $\lambda_{ep}$ to be 2.99, 2.35, 2.04, 1.84 and 1.69 for the pressures 12 GPa, 14 GPa, 16 GPa, 18 GPa and 20 GPa, respectively. All these values of the coupling constants are high [67,68] and $KB_2H_8$ is predicted to be a strongly coupled superconductor. The electron-phonon coupling constant has different representations. It can be expressed as, $\lambda_{ep} = N(E_F)V_{ep}$, where $V_{ep}$ is the electron-phonon interaction energy. In the previous section, we have found that $N(E_F)$ decreases with increasing pressure (Table 11). This implies that the decrement of $\lambda_{ep}$ with increasing pressure found by Gao et al. [19] is partly due to the pressure dependent changes in $N(E_F)$. We have calculated the superconducting transition temperatures of $KB_2H_8$ at different pressures using the widely employed McMillan equation [69] given below:

$$T_c = \frac{\theta_D}{1.45} \exp\left[-\frac{1.04(1+\lambda_{ep})}{\lambda_{ep} - \mu^*(1+0.62\lambda_{ep})}\right] \quad (15)$$

We have used the computed values of Debye temperature and Coulomb pseudopotential to estimate pressure dependent $T_C$. The values of electron-phonon coupling parameters for different pressures have been taken from a previous study [19]. The computed $T_C$ values at different pressures are presented with the relevant parameters in Table 12.

**Table 12**. The predicted superconducting transition temperatures at different pressures of cubic $KB_2H_8$.

| Pressure, P (GPa) | $\lambda_{ep}$ [19] | $\mu^*$ | $\theta_D$ (K) | $T_C$ (K) | $T_C$ (K)* [19] |
|---|---|---|---|---|---|
| 12 | 2.99 | 0.143 | 639.77 | 96.44 | 146.0 |
| 16 | 2.04 | 0.140 | 706.77 | 84.80 | 132.0 |
| 20 | 1.69 | 0.137 | 754.53 | 77.50 | 124.0 |

*Calculated with $\mu^* = 0.10$

It is noted from Table 12 that the computed values of $T_C$ shows the similar decreasing trend with increasing pressure as found previously [19]. But the values obtained here are lower than those found by Gao et al. [19]. Once source of this discrepancy is due to the use of a lower value of $\mu^*$ in Ref.[19] which was selected somewhat arbitrarily.

*3.8 Optical properties*

We have presented the optical characteristics of KB$_2$H$_8$ in this section. We have found that pressure has a little effect on various frequency/energy dependent optical parameters. Therefore, the plots are selected here only for a pressure of 12 GPa.

The dielectric functions are calculated with the following equations:

$$\varepsilon(\omega) = \varepsilon_1(\omega) + i\varepsilon_2(\omega) \tag{16}$$

Here, $\omega$ is the photon angular frequency, $\varepsilon_1(\omega)$ and $\varepsilon_2(\omega)$ are the real and imaginary parts of the dielectric functions, respectively. The imaginary part of the dielectric function can be calculated from [21,66]:

$$\varepsilon_2(\omega) = \frac{2\pi e^2}{\Omega \varepsilon_0} \sum_{k,v,c} |\psi_k^c| \boldsymbol{u}.\boldsymbol{r} |\psi_k^v|^2 \delta(E_k^c - E_k^v - E) \tag{17}$$

Here, *e* is the electronic charge, $\Omega$ is the unit cell volume, $\boldsymbol{u}$ is the unit vector along the polarization direction of the incident electric field and $\psi_k^c$ and $\psi_k^v$ are the wave functions for conduction and valence band electrons at a particular *k*, respectively. Equation 17 describes the interband optical transition of charge carriers which dominates in most of the materials. Nevertheless, due to the metallic nature of KB$_2$H$_8$, the intraband optical transitions also play a role at far infrared regions of electromagnetic wave spectrum, i.e., at low energy parts of the spectrum [70,71]. This contribution is taken into consideration by including a Drude damping term in the calculations [72,73]. Therefore, we have obtained the optical parameters with a screened plasma energy of 5 eV and a Drude damping term of 0.05 eV.

Figure 8 (a) shows the real and imaginary parts of the dielectric function with photon energy up to 10 eV. The large negative values of $\varepsilon_1(\omega)$ are observed in the far infrared region. This indicates the Drude like behavior as seen in metals, agreeing with the electronic band structure of KB$_2$H$_8$. On the other hand, it is observed that $\varepsilon_2(\omega)$ decreases sharply from a high free electron like value in the infrared region. The energy where both the parts of $\varepsilon(\omega)$ approach low values in the high energy region corresponds to the plasmon energy/frequency [74]. The real part of the dielectric function determines the polarization of charge carriers due to the incident electric field, while the imaginary part signifies dielectric loss within the material.

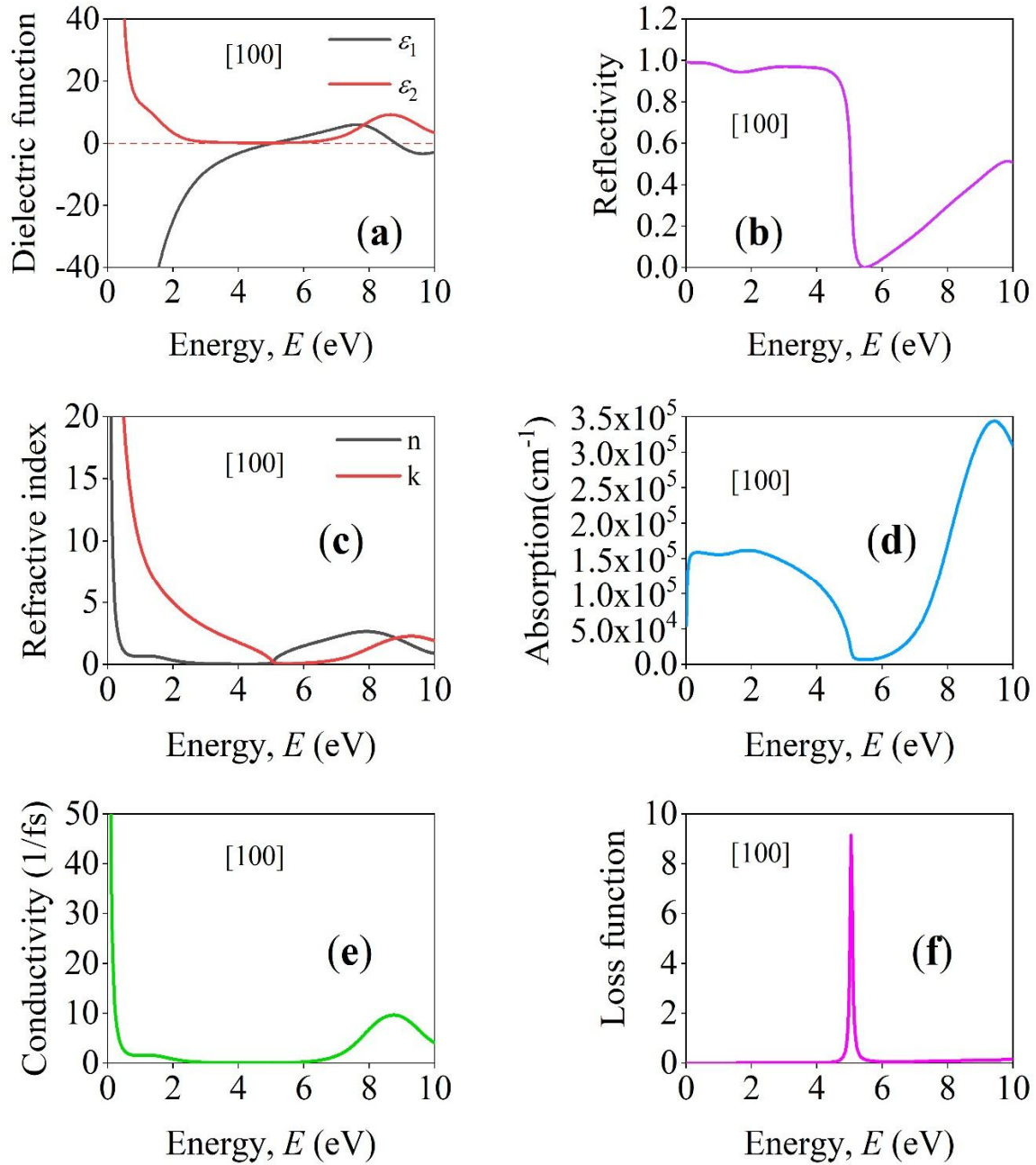

**Figure 8**: Optical properties of cubic $KB_2H_8$ at 12 GPa. The polarization direction of the incident electric field is along the [100] direction.

Fig. 8 (b) represents the reflectivity spectrum. The reflectivity is nonselective and almost 100% in the spectral range covering the infrared to near ultraviolet (UV) region. This suggests that $KB_2H_8$ is an excellent reflector of visible light [73]. Thus this compound has very high potential to be used as a solar reflector. The reflectivity falls rapidly at the energy close to the plasma frequency which is also consistent with the position of the peak in the energy loss function.

Subsequent, nearly linear, increase in the reflectivity for energies above 5.5 eV corresponds to the high TDOS peaks in the conduction bands.

The refractive index (*n*) and the extinction coefficient (*k*) are shown in Fig. 8 (c). The real part of refractive index is associated with phase velocity of light in the compound, while the imaginary part, often termed as the extinction coefficient, indicates the amount of attenuation when the electromagnetic wave traverses through the medium. The extinction coefficient is closely related to the absorption coefficient. High values of *k* in the infrared region are due to the metallic nature of $KB_2H_8$ (free electron response). The refractive index is quite high in infrared region. It also shows a peak at 8.0 eV in the UV region. High refractive index compounds can be used in optoelectronic devices such as in liquid crystal displays (LCDs), organic light emitting diodes (OLEDs) and quantum dot (QLED) televisions [75]

The absorption coefficient indicates about the optimum solar energy conversion efficiency and about how far light (of specific energy) penetrates into a material before being absorbed. The absorption spectrum for $KB_2H_8$ is shown in Fig. 8 (d). The absorption coefficient starts from 0 eV supporting the metallic nature as found in the band structure calculations. Absorption coefficient is nonselective in the infrared and visible region. It sharply decreases at higher energies and falls to zero ~ 5.5 eV. At this particular energy, the compound is expected to be completely transparent to the incident light. Further increase in the photon energy gives a high-value peak in the absorption coefficient at ~ 9.0 eV in the UV region. Close to this energy, $KB_2H_8$ becomes an efficient absorber of UV radiation. These results indicate that $KB_2H_8$ is a promising absorbing material for visible and the UV light.

The real part of optical conductivity of $KB_2H_8$ is shown in Fig. 8(e). The photoconductivity also starts at zero photon energy complementing the result of metallic character. The position of the high energy peak in the optical conductivity coincides with that in the absorption coefficient indicating that the UV photons of this particular energy (~ 9.0 eV) contributes significantly to the enhancement of the optical conductivity.

The energy loss spectrum is given in Fig. 8(f). This parameter is important in the dielectric formalism and is used to understand the screened optical excitation spectra. The highest peak of the energy loss spectrum appears at a particular incident light energy and gives the information about the bulk plasma frequency/energy. The plasma energy is found to be 5.0 eV for cubic $KB_2H_8$. A material becomes transparent when the energy/frequency of the incident light is higher than its plasma energy/frequency. Usually, above the plasma energy the optical response of a material becomes similar to that of an insulator.

## 4. Conclusions

In this study, we have studied the structural, elastic, mechanical, thermophysical, electronic, superconducting and optical properties of cubic $KB_2H_8$ under different uniform pressures. Most of the results presented are novel. The compound under investigation was found to be thermodynamically stable for pressures at and above 8 GPa. The compound is also elastically stable from 8 GPa to 24 GPa. The compound $KB_2H_8$ in the stable state is ductile in nature. It is also highly machinable. The levels of machinability and dry lubricity increase with rising pressure. The cubic $KB_2H_8$ is relatively soft in nature with low level of elastic anisotropy. The values of Debye temperature is high consistent with sound velocity; both these parameters increase with increasing pressure. The Grüneisen parameter is high suggesting that lattice anharmonicity is significant in $KB_2H_8$. The compound also possesses high minimum phonon thermal conductivity. The electronic band structure reveals clear metallic behavior. Metallicity originates from the hybridized H-*s* and B-*p* electronic orbitals. The optical parameters show close correspondence with the electronic band structure. The compound under study is an extremely efficient reflector of infrared and visible light. The refractive index is high in the low energy region. The compound is an efficient absorber of UV radiation. The electronic band structure close to the Fermi level becomes more dispersive with increasing pressure; as a result, the electronic energy density of states is lowered as pressure increases. This reduces both the electron-phonon coupling constant and the repulsive Coulomb pseudopotential. The decrement is more prominent for the electron-phonon coupling constant and therefore, the superconducting critical temperature of cubic $KB_2H_8$ decreases with increasing pressure.

**Acknowledgements**
S. H. N. acknowledges the research grant (1151/5/52/RU/Science-07/19-20) from the Faculty of Science, University of Rajshahi, Bangladesh, which partly supported this work. Md. A. A. acknowledges the financial support from the Bangabandhu Science and Technology Fellowship Trust for his Ph.D. research.

**Data availability**

The data sets generated and/or analyzed in this study are available from the corresponding author on reasonable request.

**Declaration of interest**

The authors declare that they have no known competing financial interests or personal relationships that could have appeared to influence the work reported in this paper.

# CRediT author statement

**Md. Ashraful Alam:** Methodology, Software, Formal analysis, Writing-Original draft. **F. Parvin:** Supervision, Writing-Reviewing and Editing. **S. H. Naqib:** Conceptualization, Supervision, Formal analysis, Writing- Reviewing and Editing.